\documentclass{kapproc} 

\usepackage{procps} 

\usepackage[dvips]{graphicx}

\upperandlowercase

\kluwerbib 

\def\aj{AJ}%
\def\apj{ApJ}%
\def\apjs{ApJS}%
\begin{document}


\articletitle{The Opacity of Spiral Galaxy Disks}
\articlesubtitle{Dust opacity from calibrated counts of distant galaxies.}

\author{Benne W. Holwerda\altaffilmark{1,2}, R.A. Gonz$\rm \acute{a}$lez\altaffilmark{3}, Ronald J. Allen\altaffilmark{2} and P.C. van der Kruit\altaffilmark{1}}
\email{holwerda@stsci.edu}


\altaffiltext{1}{Kapteyn Astronomical Institute, Postbus 800, 9700 AV Groningen, The Netherlands}
\altaffiltext{2}{Space Telescope Science Institute, 3700 San Martin Drive, 21218 MD Baltimore, USA}
\altaffiltext{3}{Centro de Radiastronom$\rm \acute{\imath}$a y Astrof$\rm \acute{\imath}$sica, Universidad Nacional Aut$\rm \it \acute{o}$noma de M$\rm \acute{e}$xico, 58190 Morelia, Michoac$\rm \acute{a}$n, Mexico}

\begin{abstract}
The opacity of foreground spiral disks can be probed from the number 
of distant galaxies seen through them. To calibrate this number for effects 
other than the dust extinction, Gonzalez et al (1998) developed the
"Synthetic Field Method". A synthetic field is an extincted Hubble Deep 
Field added to the science field. The relation between the dimming and 
the number of retrieved synthetic galaxies calibrates the number found
in the science field. Here I present results from counts in 32 HST/WFPC2 
fields. The relation between opacity and radius, arm and disk, surface 
brightness and HI are presented. The opacity is found to
be caused by a clumpy distribution of clouds in the disk. The brighter parts 
of the disk -the center and arms- are also the more opaque ones. The dust 
distribution in spiral disks is found to be more extended than the stellar disk. 
A comparison between HI column densities and opacity shows little relation 
between the two.
\end{abstract}


\section{Introduction}

The measure and extent of the dust absorption by spiral disks has been the
subject of study and even controversy for some time. 
New models of the disk's energy household (e.g. \cite{Dopita}) 
and a wealth of observational data (e.g. \cite{Regan} and \cite{Kennicutt}), promise a better understanding of the role of dust in the disks. 
Still, there are several questions regarding the distribution of dust clouds in the 
spiral disk which should be addressed: to which radius does dust extend and is 
dust spatially correlated to the stellar distribution or the atomic hydrogen gas?

One possible technique to answer these questions is to use the number of distant 
galaxies seen through a foreground disk as an indicator of disk opacity. 


\begin{figure}[ht]

\includegraphics[width=8.5cm]{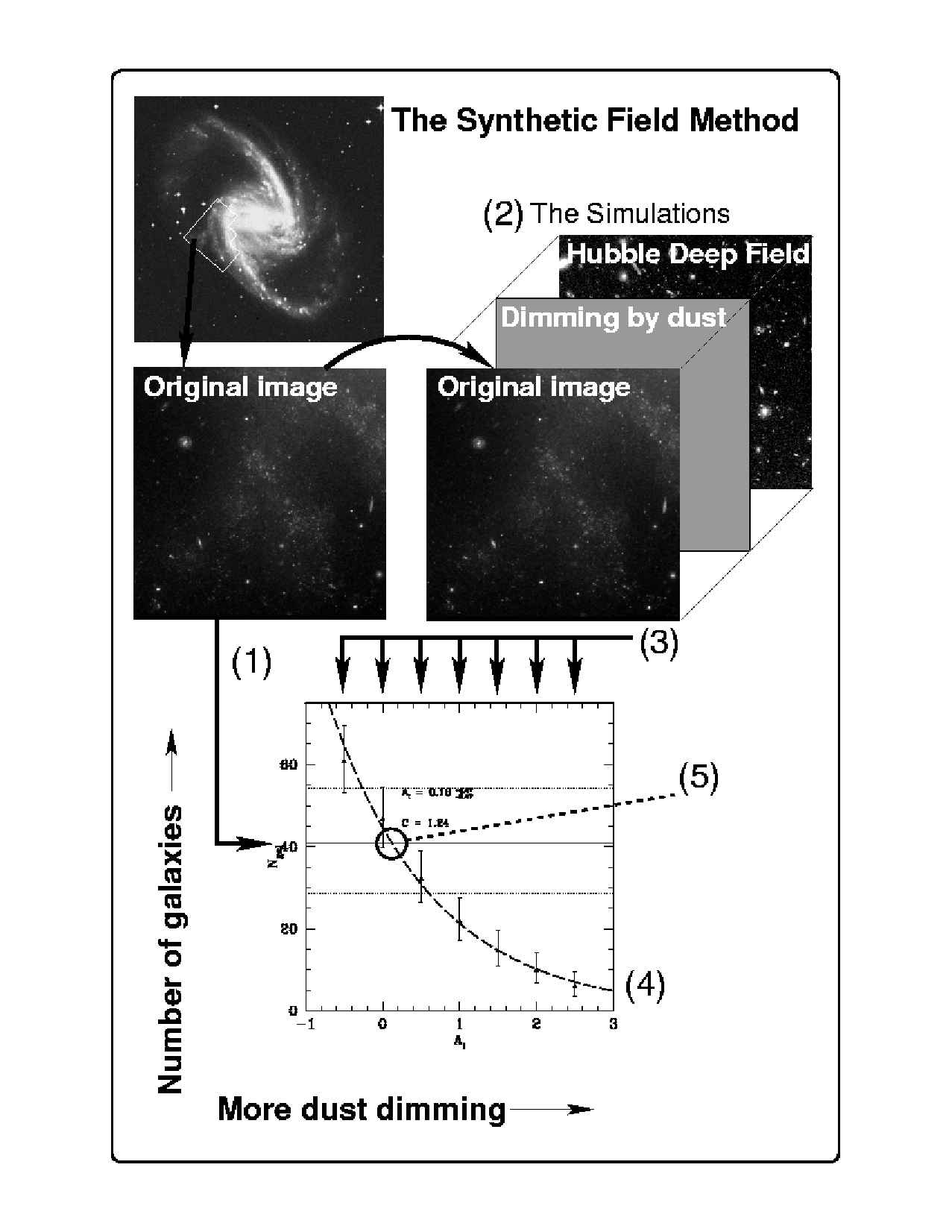}
\narrowcaption{A schematic of the ``Synthetic Field Method''. First a WFPC2 field is retrieved 
from the Hubble Space Telescope archive and redrizzled. \protect\\ 
The SFM steps are:\protect\\ 
1. The number of distant galaxies in the original science field are counted. \protect\\ 
2. The ``synthetic fields'' are made by combining a dimmed Hubble Deep Field 
with the science field.\protect\\ 
3. The numbers of synthetic galaxies are counted in the synthetic fields.\protect\\ 
4. $\rm A = -2.5 ~ Log_{10}(N/N_0)$ it fitted to the number of synthetic galaxies ($N$) as a function 
of the applied dimming ($A$).\protect\\ 
5. From the intersection between the number galaxies in the science field and 
the fit, the average dimming in the image is found. }
\label{fig:method}
\end{figure}

\section{What is the ``Synthetic Field Method''?}

For a useful measurement, one needs to calibrate the number of distant galaxies 
seen through a foreground for effects other than the extinction by dust. 
The calibration is done with the ``Synthetic Field Method'' (Figure \ref{fig:method}).
The distant galaxies in the science field are counted (step 1). An average 
background field is added to the science field to create a ``synthetic field''. 
The average background field is artificially dimmed in progressive ``synthetic fields'' 
(step 2). The numbers of retrieved synthetic galaxies for each value of the dimming 
is found (step 3). The relation $A = ~ -2.5 ~ Log_{10}(N/N_0)$ is fitted to the relation 
between the numbers of galaxies (N) and the dimming (A) of the synthetic fields 
(step 4). The free parameters are the slope ($C$) and the number of galaxies 
without any dimming ($N_0$). The intersection between the fit and the number 
of actual distant galaxies found in the science field gives the average opacity of 
the science field (step 5)\footnote{A comprehensive description of the ``Synthetic 
Field Method'' and the uncertainties involved is presented in \cite{Holwerda05a} 
and \cite{Holwerda05}. The effects of disk's characteristics on the accuracy is 
discussed in \cite{Holwerda05d}.}. This comparison between counts can be done for individual 
fields but also for multiple fields, combined based on disk characteristics: 
radius, arm or disk region, surface brightness or HI column density. 
\footnote{The numbers of distant galaxies in a single HST/WFPC2 field are only sufficient for 
a relatively uncertain opacity measurement. To counter this much more solid angle 
from several HST observations need to be combined.}

\section{The opacity profile of spiral disks}

\begin{figure}[ht]
\includegraphics[width=8cm]{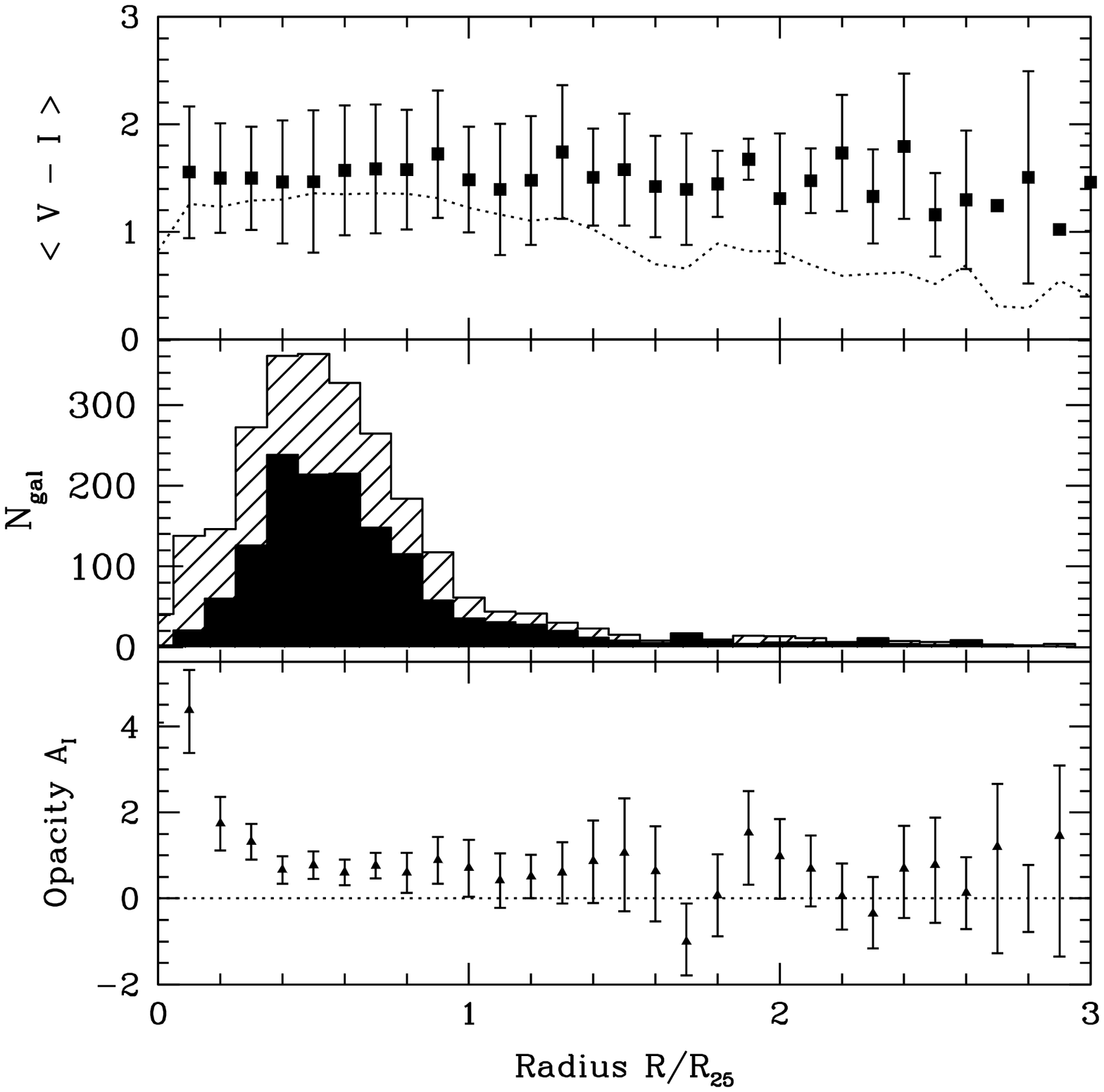}
\narrowcaption{The stacked radial opacity profile for all 32 galaxies in the sample. 
The counts of galaxies were ordered according to deprojected radial distance from 
the foreground galaxy's center, expressed in $R_{25}$. 
Top: the average color of distant galaxies in the science fields. Middle: the number 
of distant galaxies in the science field (filled) and the synthetic fields without any 
dimming (shaded). Bottom: the disk opacity as a function of radius derived from the 
numbers of distant galaxies. Since most of the WFPC2 fields were pointed at the disk 
of these galaxies, the opacity measurements are the most accurate there (from \cite{Holwerda05b}).}
\label{fig:RA}
\end{figure}

Figure \ref{fig:RA} shows the stacked radial opacity profile of our entire sample 
of 32 HST/WFPC2 fields. 
From Figure \ref{fig:RA}, we can conclude that the average 
radial opacity profile of spiral galaxies remains relatively flat for most of the disk and 
only tapers off somewhere beyond the $R_{25}$. This was already suspected from sub-mm observations (e.g. \cite{Nelson98} and \cite{Zaritsky}). This profile has not 
been corrected for inclination effects, as this depends greatly on the assumed dust geometry.

The average color of the galaxies in the science fields does not change however 
(Figure \ref{fig:RA}, top panel). If the diminishing effect of the dust in the ISM would have 
been due to a uniform screen, the average color would have changed with the total amount 
of attenuation according to an Extinction Law. However, if the dust in the disk is relatively 
patchy, the attenuation is derived from the number of absent galaxies while the color is 
from those that have been detected relatively unhindered by dust in the disk. The average 
disk opacity in Figure \ref{fig:RA} can therefore be interpreted as a cloud covering factor. 

\begin{figure}[ht]
\includegraphics[width=7cm]{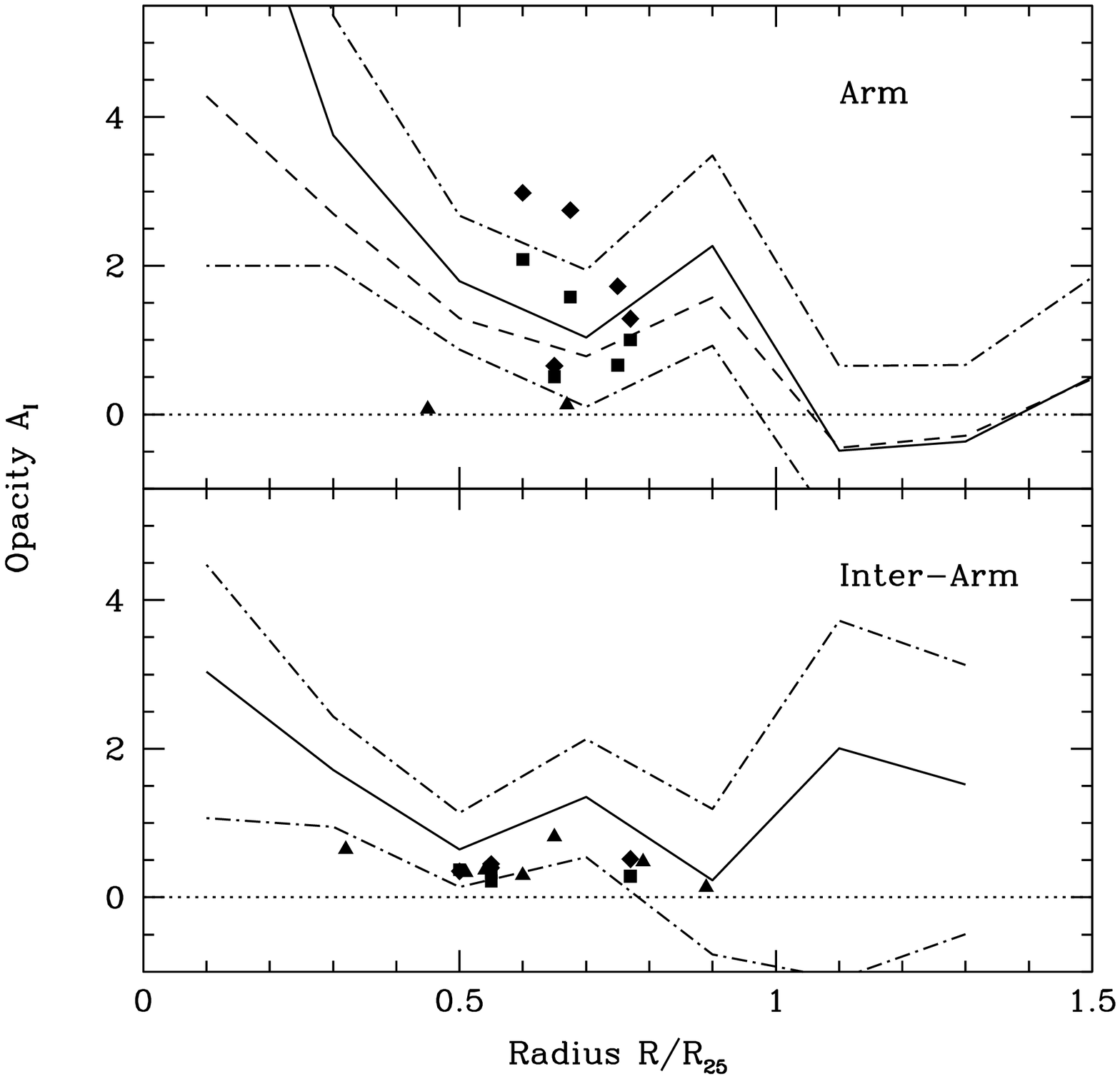}
\narrowcaption{The stacked radial opacity profiles for the arm and inter-arm regions of 
the disks. The filled squares and triangles are the measurements by \cite{kw00a} and 
\cite{kw00b} respectively. The dashed/dotted lines are the uncertainties in the SFM 
measurements. higher values for spiral arm opacity have been found in earlier photometric studies (e.g. \cite{Elmegreen80, Block94b,Peletier95,Kuchinski98}) }
\label{fig:KW}
\end{figure}

Separate profiles for the arm and disk sections of the foreground galaxy can also be 
made (Figure \ref{fig:KW}). These opacity profiles show a strong radial relation in the 
arms and a much flatter profile in the rest of the disk. When these profiles are compared 
to the opacity values from occulted galaxy pairs (\cite{kw00a,kw00b}), there is a 
remarkably good agreement between these completely different methods. \cite{kw00a} 
also found a gray behaviour but \cite{kw01a} found that the Galactic Extinction law 
re-emerged for measurements below a linear scale in the disk smaller than 100 pc. 

\section{Dust, gas and stars}

The strong radial relation for the spiral arms hints at a direct relation between 
the amount of dust in the disk and the surface brightness of the disk. A similar 
relation between overall disk opacity and brightness was found by 
\cite{Tully98,Masters03}. 
However the much flatter distribution of the opacity in the rest of the disk does 
suggest that the dust may be more distributed like the atomic hydrogen.

\begin{figure}[ht]
\includegraphics[width=7cm]{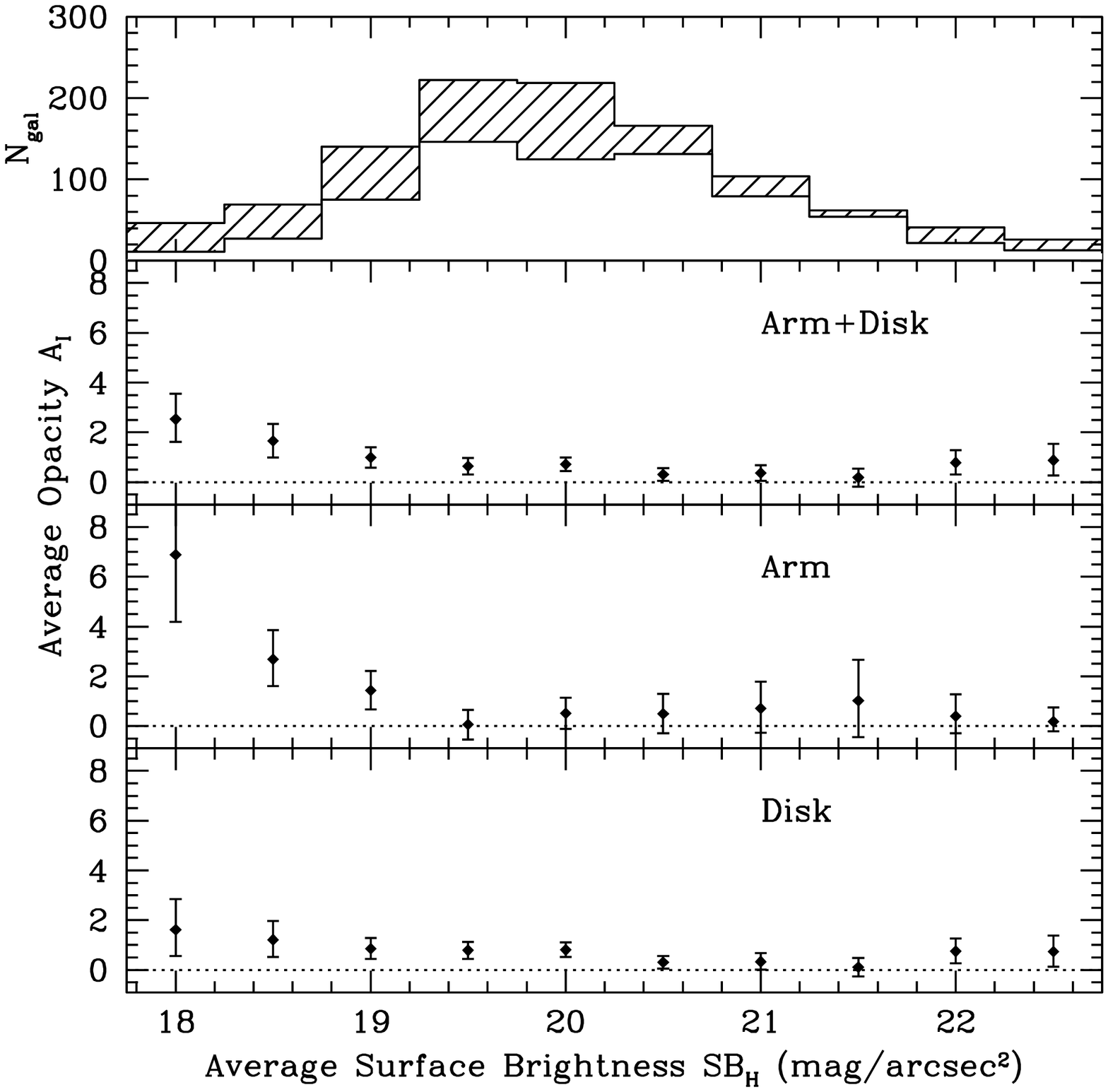}
\narrowcaption{The disk opacity as a function of surface brightness. The distant 
galaxies -real and synthetic- are sorted by surface brightness in the entire 
sample. Top: the histograms of distant galaxies (shaded for the synthetic 
fields and the line for the science field). Top middle: the derived opacity as 
a function of surface brightness. Bottom two plots: the surface-brightness 
and opacity plots for the arm (top) and disk (bottom) parts of the spiral galaxy.(from \cite{holwerda05e})}
\label{fig:SB}
\end{figure}

To find the relation between opacity and surface brightness, each distant galaxy, 
from both synthetic and science fields, are flagged with the surface brightness 
in 2MASS images. The numbers of galaxies as a function of near-infrared 
surface brightness are shown in Figure \ref{fig:SB} for all of the disk and the 
arm and non-arm regions separate. As could be expected from Figure \ref{fig:KW}, 
there is a strong relation between surface brightness and opacity in the arm 
regions but none in the non-arm part of the disk.

\begin{figure}[ht]
\includegraphics[width=7cm]{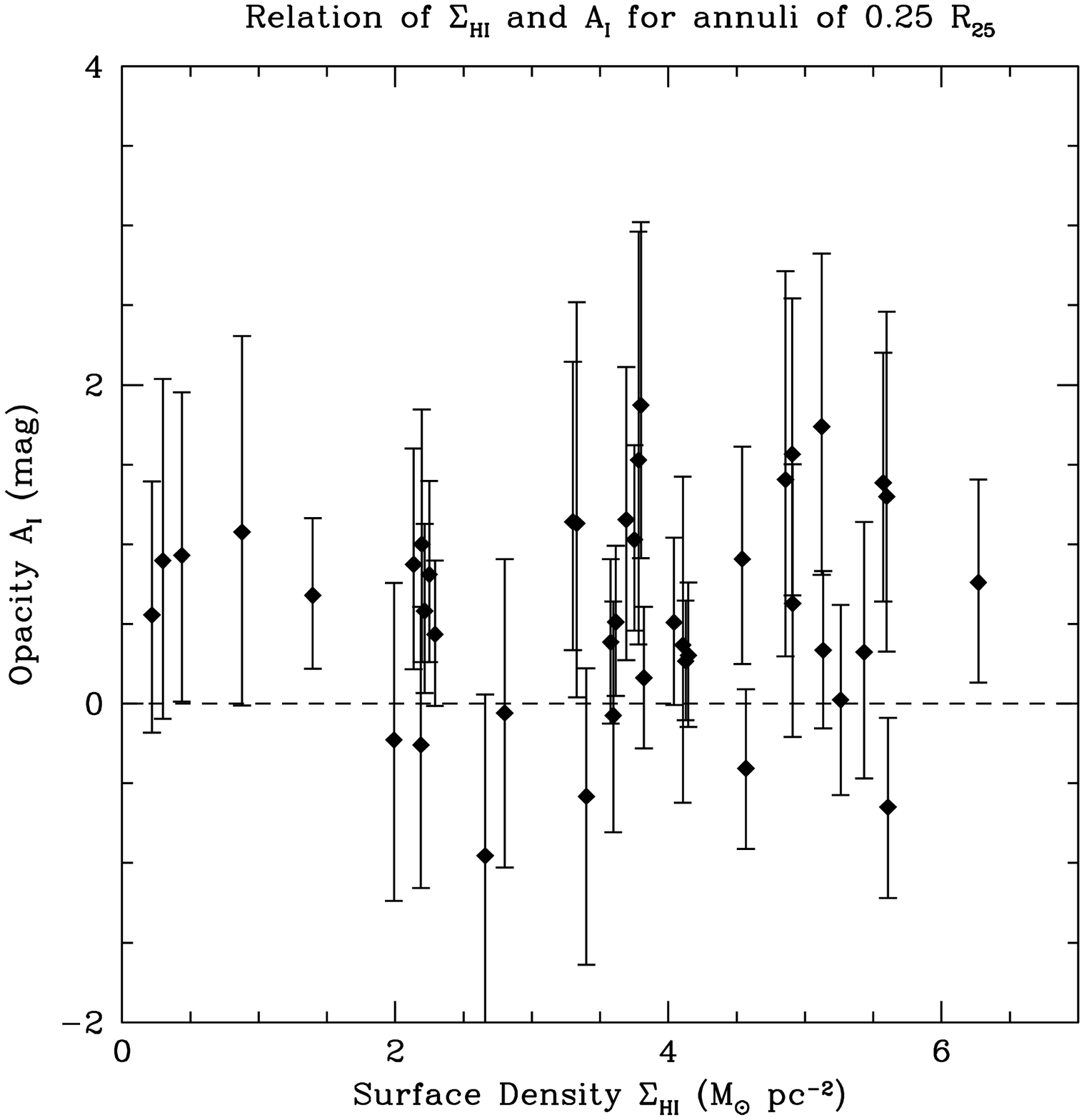}
\narrowcaption{The ratio of HI column density and opacity in radial annuli of 0.25 $R_{25}$. Opacity measurements in individual galaxies remain uncertain due to cosmic variance in the number of distant galaxies behind each foreground galaxy. An improvement would include compare galaxy number directly as a function of HI column density (e.g. \cite{Cuillandre01}), much like Figure \ref{fig:SB}. (from \cite{Holwerda05c})}
\label{fig:HI}
\end{figure}

The relation between atomic hydrogen and dust might reveal more on the
hidden constituents of the ISM. Due to limitations of the SFM measurements 
and lack of HI maps, Figure \ref{fig:HI} shows the ratio of the HI column density from
radial profiles from the literature and the opacity measured from counts of galaxies 
in the same radial bin. No direct relation is evident, mainly because opacity profiles rise towards the center of galaxies and HI profiles stay flat or decrease. It is possible that dust correlates with
the total gas content in the disk, both the molecular and the atomic. 

\section{Final Remarks}

The counts of distant galaxies have proven themselves to be a good indicator of disk 
opacity, provided they are calibrated with the SFM. Improved statistics from many HST 
fields have lent insight into the distribution of dust in the disk and into the relation with 
the stellar content and atomic gas. The radial opacity profile extends to well beyond the 
optical disk and appears caused by patchy dust. The dust clouds correlate with stellar 
light in the arms but there is additional constant disk component which is more dominant 
at higher radii.

The radial profile implies that there are dust clouds outside the optical disk. This seems in 
agreement with the SINGS extinction maps (\cite{Regan}) and PAH emission maps from 
the SINGS project (\cite{Kennicutt}) but an exact comparison remains to be made. However, 
we know that starformation and HII regions can be found at hitherto unprecedented radii. 
The presence of dust is therefore not unreasonable. The question remains how much there is.

One implication of the relation between opacity and surface brightness in the arms 
is that a light-profile derived from the non-arm regions can be a maximum disk or a 
scaling of that but that a profile that includes the arm regions will be less concentrated 
than the actual distribution of mass. This may be another contributor to a slightly different 
normalization of the \cite{BelldeJong} color-M/L relation (\cite{deJong}).

The profiles of opacity and HI column density are not correlated. Since the opacity keeps 
rising in the center, it is very possible that the overall disk opacity relates in some way to 
the total gas content or even just the molecular component. 

A similar radial profile and HI-to-dust ratio for spiral disks have also been found in sub-mm observations of disks. This implies that a substantial fraction of the dust responsible for the disk opacity is cold. 

Future application of the counts of distant galaxies include probing the edge of the dust disk, constraining the average cloud size, a comparison with sub-mm observations to constrain dust emissivity and a direct comparison with column density 


\end{document}